\documentclass[12pt]{article}
\usepackage{graphicx}
\usepackage{epsfig,a4}

\begin{document}

\markboth{Abramovsky V.A.,Dmitriev A.V}{Possible enhancments of higgs trrigger}
\title{Possible improvements of higgs trigger at LHC}
\author{\footnotesize V.A.ABRAMOVSKY, A.V. DMITRIEV \\
\footnotesize Novgorod State University, B. S.-Peterburgskaya Street 41,\\
\footnotesize Novgorod the Great, Russia,\\
\footnotesize 173003}
\maketitle

\begin{abstract}
In this work we consider two ways to improve signal to background cross-sections ratio for higgs searchings at LHC: likelyhood method and advanced rapidity trigger. Both methods are universal enough, likelyhood method can be applied to any processes with many observables, and advanced rapidity trigger can separate any colourless scattering processes and processes with colour charge at $t$ channel.

\end{abstract}

\section*{Introduction}

Searching of the higgs boson in one of the main purpose of LHC program. At  SM and MSSM higgs will be observable, but at the limits of detectors capacities. If the theory at TeV scale is differ than SM or MSSM, we risk to fail to observe higgs. From the other side, we want to  observe most, and rare too, decaying channels of higgs. Common methods detect only leading channels (at given $M_H$) of higgs decay. So, it is very important to enhance higgs detection methods.

Higgs searchings at LHC based on Standart (Minimal Supersimmetric) Model, which state properties of Higgs at given higgs mass in SM and on mass and $tan(\beta)$ in MSSM. LEP results excluded some part of $M_H,tan(\beta)$ plane, lower limit for higgs mass is $M_H>100GeV$. From other side, unitarity gives upper limit $M_H<1TeV$.

At this mass range there are two main channels of higgs production, gluon-gluon fusion $gg \rightarrow H$ and vector boson fusion (VBF) $WW \rightarrow H$. Leading channel is gluon-gluon fusion, but VBF can be easily separated from background, because of VBF generates two energetic jets at hight rapidities and rapidity gap in central region. Rapidity gaps also exist in pomeron-fusion higgs produstion which can be considered as color screening in gluon-gluon fusion by additional gluon exchange. In this work we consider only VBF channel.

Common method for separating signal from background is cut-by-cut method. Good review of applicability of this method for higgs searching at LHC is in Ref.\cite{review_higgs}. More detailed considering of $H\rightarrow\tau\tau$ is in Refs.\cite{atl_phys_2002_018},\cite{atl_phys_2003_004}. Extensions of traditional technic are usually based on modifications or adding of cuts (see Ref.\cite{Mellado:2004tj}).
Multivariate technique is out of mainstream for higgs searchings at LHC, because of neural networking results is not stable and can not be controlled.

At present time, ATLAS higgs dtector (see.Ref\cite{review_higgs}) have good signal-to-background ratio, but most  ($\sim 90\%$) of the higgs events are rejected. We purpose two way to improve this situation. At first, we re-analize applicability of the likelyhood analisys to the higgs searching. At second, we suggest additional quantity to select signal events. Both improvements are complementary and nicely connected.

In this paper we investigate the vector boson fusion channel of the higgs production \cite{Bjorken},\cite{WW} mainly. In Sec.\ref{Likelyhood} we reanalyse multivariate technics and check applicability to $H\rightarrow\tau\tau$ channel.

\section*{Likelyhood method}
\label{Likelyhood}

Traditionally higgs events is selected by applying of the on-by-one cuts. Without cuts, we have vanishing signal to background ratio (see Table 1, taken from [\cite{jet1}]). Each next cut refuse some part of events, and cuts is selected to remove most of the backround and save most of the signal events. At the end of the process, we have very good signal to background ratio, but signal is reduced by some order of magnitude. 
Common cuts is requrement of two jets with high rapidity, veto on the jet activity in the central region and existence of the higgs decaying products. Also there is many other observables with distributions differs from signal to background events. Any cut stand some pair of observable and region, and event is passed throw cut only if observable is in this region.

Following cuts method, we risk  (and actually do) to reject many of the signal events. This happens, than one cut reject event, but other signs clearly show, that this event is signal.

So, we suggest to reconstract the trigger mechanism. 
At first, let`s simplify problem by assumption, that there is no interference between signal and background processes. Than our problem can be easily formulated as the task to calculate probability for any  event to be produced by signal or by background process. Let`s define $P(Signal|X)$ and  $P(Background|X)$ as the probability at given $X$, that this $X$ is produced by signal and  background process, respectively, and normalization states
\begin{equation}
P(Signal|X)+P(Background|X)=1
\end{equation}
As input, we have probabilities $P(X|Signal)$ and $P(X|Background)$ to produce event X by signal and background processes, respectively, and absolute probabilities $P(Signal)$ and $P(Background)$ with normalization conditions
\begin{equation}
\begin{array}{l}
\sum_{X}^{}{P(X|Signal)}=1\\
\sum_{X}^{}{P(X|Background)}=1\\
P(Signal)+P(Background)=1
\end{array}
\end{equation}
Then $P(Signal|X)$ can be easily calculated
\begin{equation}
P(Signal|X)=\frac{P(X|Signal)P(Signal)}{P(X|Signal)P(Signal)+P(X|Background)P(Background)}
\label{SX}
\end{equation}
If we assume, that distributions $P(X_i|Signal)$ of observables $X_i$ is independent from each other, then we can use this distributions directly to calculate $P(Signal|X)$ using (\ref{SX}) and relation
\begin{equation}
P(X|Signal)=\mathbf{\Pi}P(X_i|Signal)
\end{equation}
To use derived equation (\ref{SX}) as a trigger, we must choose lower limit of probability $P_0$. If $P(Signal|X)<P_0$, then event rejected, otherwise event accepted. This level $P_0$ must be chosen to maximaize confidence level $S/\sqrt{B}$, there $S$ is number of the signal events, and $B$ is the number of background events. From definitions,
\begin{equation}
\begin{array}{l}
S=N\sum_{X:P(Signal|X)>P_0}^{}P(X|Signal)\\
B=N\sum_{X:P(Signal|X)>P_0}^{}P(X|Background)
\end{array}
\end{equation}
there $N$ is integrated luminosity.

\begin{figure}[th]
\centerline{\psfig{file=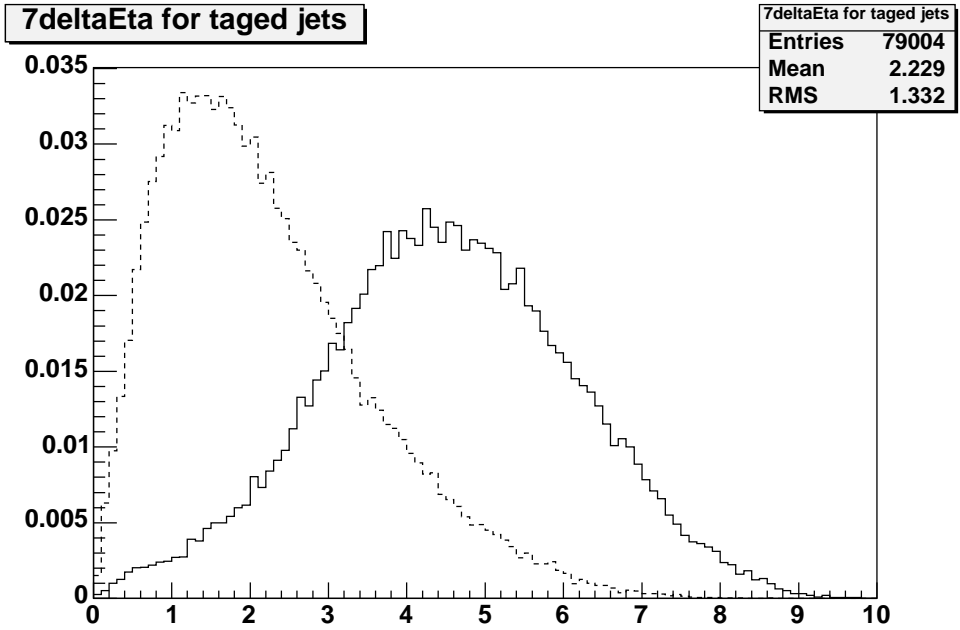,width=5in}}
\vspace*{0pt}
\caption{Rapidity distribution between leading hadron jets for VBF higgs production (solid line) and for  $t\bar{t}$ background (dashed line).}
\label{fig:example1}
\end{figure}

Advantages of multivariate method can be simplify esimated. Let`s consider a simle model, there probabilities $P_{signal}$ and $P_{background}$ had the same gaussian form width and centers located at opposite corners:
\begin{equation}
\begin{array}{l}
P(X|Signal)=\prod_i \sqrt{2/\pi} e^{-(x_i-0.5)^2} \\
P(X|Background)=\prod_i \sqrt{2/\pi} e^{-(x_i+0.5)^2}
\end{array}
\end{equation} 
and, say, 
\begin{equation}
\begin{array}{l}
\sigma_{Signal}=10 fb \\
\sigma_{Background}=10^5 fb
\end{array}
\end{equation}
at integrated luminosity $10 fb^{-1}$.
This distributions and parameters is close to higgs searches at LHC, see Fig.\ref{fig:example1}.

\begin{figure}
\includegraphics[scale=0.35]{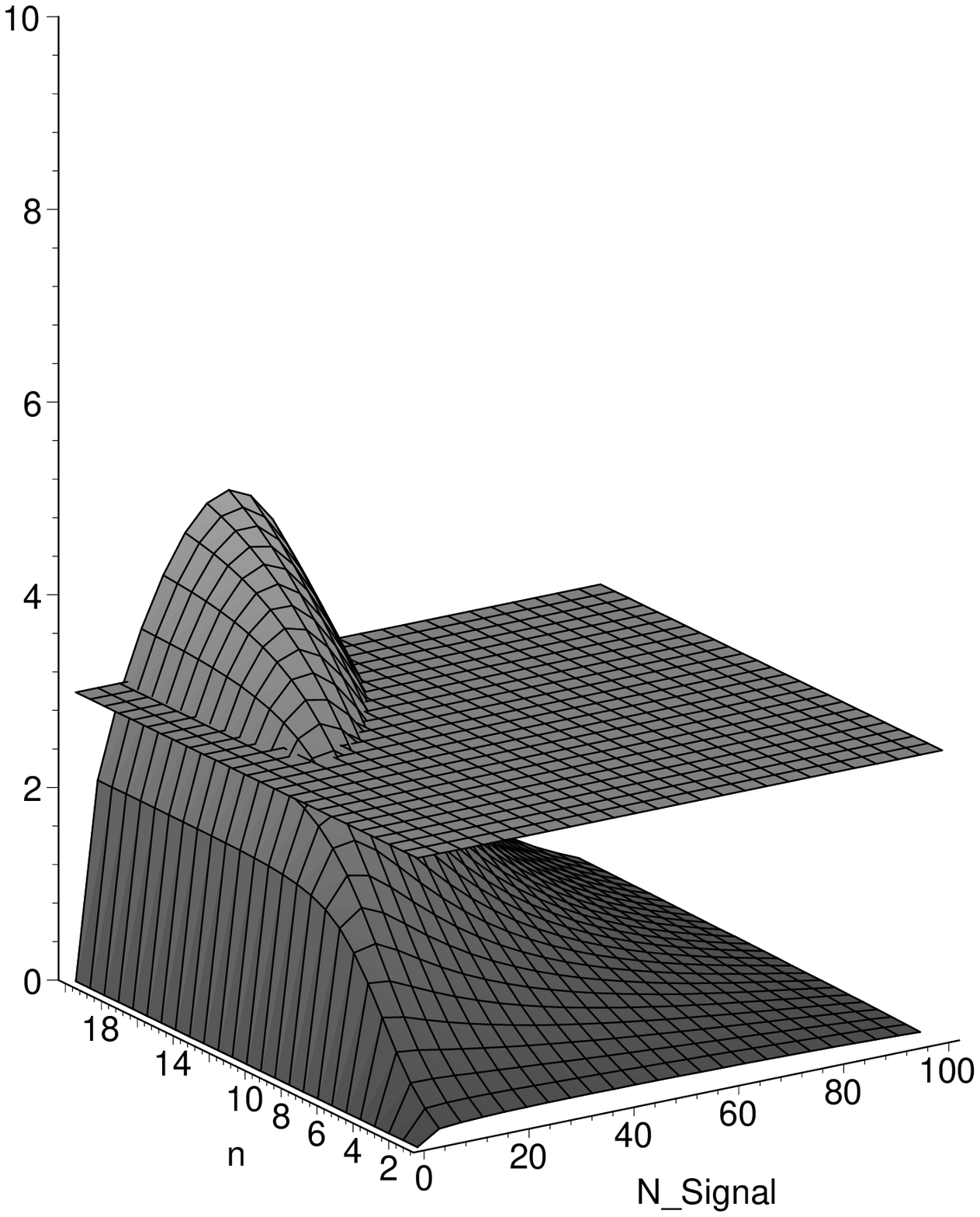}
\includegraphics[scale=0.35]{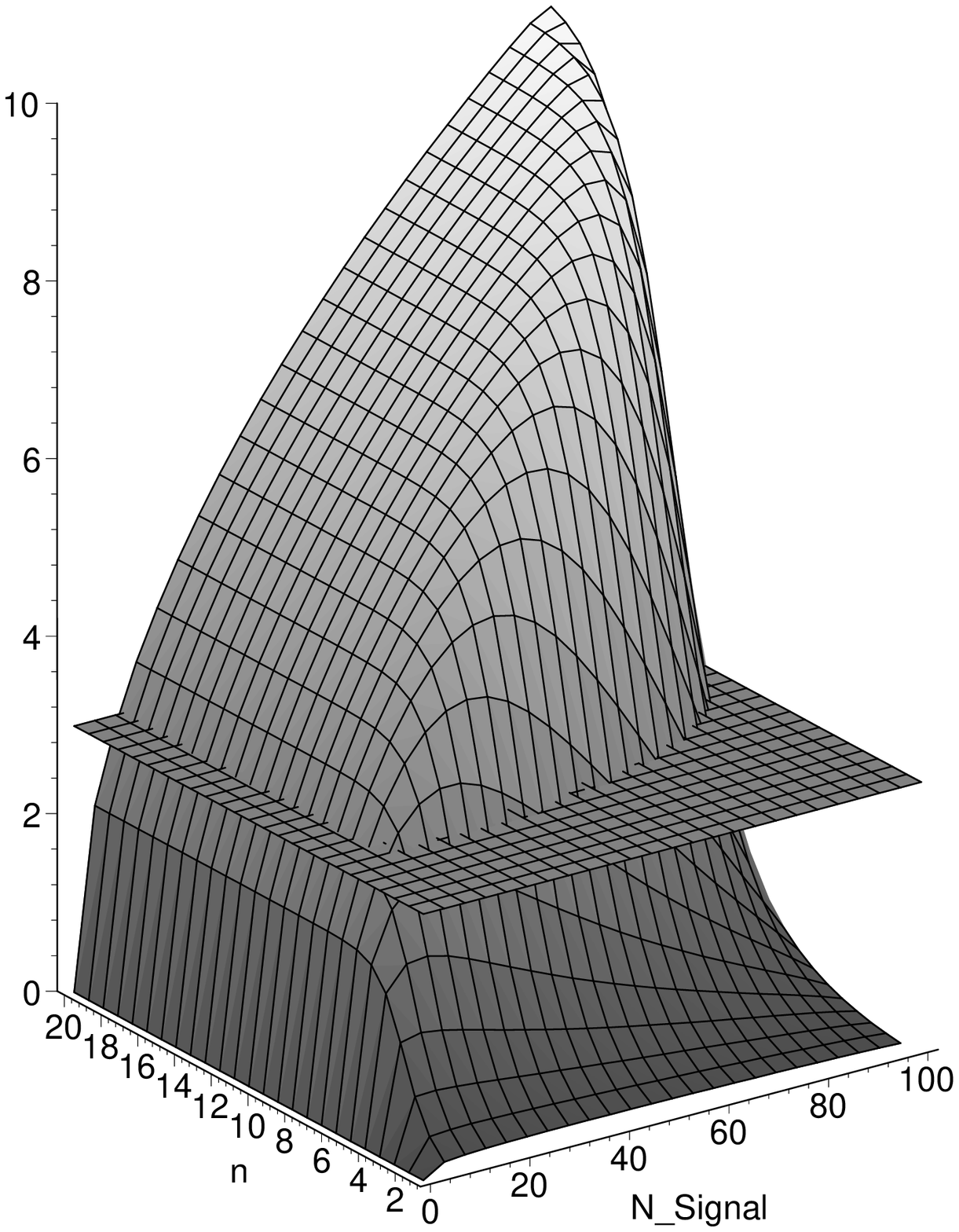}
\caption{Dependence of significance on the $N_{signal}$ - number of signal events passed throw trigger and on the $n$ - number of event characteristics. Left plot is for cut-by-cut method, right one is for multivariate method. Horizontal plane shows $3\sigma$ level.}
\label{fig:fig_3Dcompare_meth}
\end{figure}

Results for this model are presented at Fig.\ref{fig:fig_3Dcompare_meth}. It is clearly see, that if we choose level $3\sigma$ for discover, then cut-by-cut method will work only for large number of variables and trigger will accept only small part of signal events. Multivariate trigger will work on the most part of the phase space.

Efficience of multivariate trigger was checked for $H\rightarrow\tau\tau$ channel \cite{Mellado:2004tj}. Because the technical reasons, only independent variables variant was considered. As compared with cut-by-cut method (significance 6.5 units), multivariate trigger gives sigificance about 7.5 at the same number of events characteristics. Additional events characteristics will increase significance at some more units. This result is far from asymptotic estimations for indipandent variables, so correlations between events characteristics must be considered.
\newpage

\section*{Central jet veto improvements}

Let us consider  the  abstract model for processes which is going throw the fusion of colorless objects, which results to gaps if there is no re-scattering, and some sign object, independent of gap. The last one can be high  momentum $p_t$ jet or system of rare particles or something else, which can be detected independently of gap.

\begin{figure}[th]
\centerline{\psfig{file=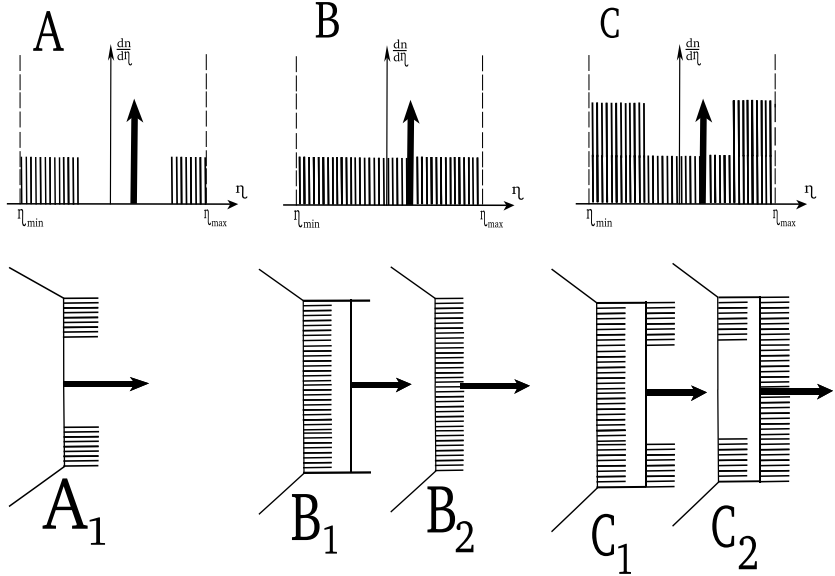,width=3in,angle=90}}
\vspace*{0pt}
\caption{Pseudo-rapidity distributions and corresponding generic processes.}
\label{fig:figABC}
\end{figure}

Diagram of such process in the case of the absence of re-scattering is marked as $A_1$ on Fig.\ref{fig:figABC}. Corresponding  pseudo-rapidity distribution of particles is marked as $A$. Bold arrow marks signal object. 

In the case of soft re-scattering (diagram $B_1$ on Fig.1) produced pseudo-rapidity distribution of particles has no rapidity gap which is usually distinguishing interesting process from the processes of the type $B_2$. The process $B_2$ have no physical interest by our assumption. So, soft re-scattering fill the signal gap, and probability of the such suppression is high, from optimistic estimation $0.85$ to pessimistic $0.99$ (see Ref.\cite{SGP1} for details). Another source of suppression is pile-up events with  more than one inelastic interaction occurs in one bunch-on-bunch collision.

This suppresion leads us to 'central jet veto' and 'forward tagging' methods, there the absence of hard jets in the central region and two hard jets on the both side of the higgs are needed to confirm colourless exchange (see.Ref\cite{review_higgs}). This traditional cuts greatly  suppress background events, background becomes $\sim 600$ time smaller, but we loose most, about $2/3$, of the signal higgs events (see Ref.\cite{review_higgs}). So, we will try to improve this situation.

We can reformulate  peculiarity of the first (signal) process to the form, that there is two 'humps' on the plateau. This peculiarity is not suppressed by soft re-scattering, because pomeron cuts produce plateau-like distributions on the pseudo-rapidity diagrams (this fact is not trivial, but it is well experimentally tested). So, after re-scattering we see two 'humps' on plateau again, but plateau is up by pomeron differential multiplicity $\frac{dn}{d\eta}$ depending on $\sqrt{s}$, but not on $\eta$.

We purpose to examine processes producing the pseudo-rapidity distribution $C$ on Fig.1, where there are two 'humps' on the both sides and plateau containing signal object. This situation is differ from the situation $B$, because we know, that some gap is produced. Generic processes is divided to two classes. At first, we have process $C_1$, containing colorless produced signal object, and $C_2$ where signal object is produced by color states (usually gluons). Inclusive production in $C_1$ of the signal object is more probable, than  exclusive one in $B_1$, but process $C_2$ is less probable, than $B_1$. So, if situation $B$ is usually produced by color production of signal object, situation $C$ is more probably produced by the colorless production of the signal object, and we can derive interesting results from this difference.

The same arguments is applicable in the case of pile-up events, we have only to cut re-scattering diagrams $B_1$, $C_1$, $C_2$ to get two independent events in each case.

Before we go to the realistic constructions, let`s consider the simple model with assumption, that all processes are factorizable. 

Let`s calculate the probabilities of producing pseudo-rapidity distributions shown at Fig.1, at given impact parameter of interaction:
\begin{equation}
P_A=P^{SO}_{white}(1-P_{inelastic})
\label{eqA}
\end{equation}
\begin{equation}
P_B=P^{SO}_{color}+P^{SO}_{white}P_{inelastic}
\label{eqB}
\end{equation}
\begin{equation}
P_C=P^{SO}_{color}P_{DD}+KP^{SO}_{white}P_{inelastic}
\label{eqC}
\end{equation}

Here $P^{SO}_{color}$ is probability of making signal object by fusion of two color objects, $P^{SO}_{white}$ is the exclusive one by fusion of two colorless objects, $P_{inelastic}$ is the probability of inelastic re-scattering (survival gap probability is $P_{SGP}=1-P_{inelastic}$ at given $b$), $P_{DD}$ is the probability of double diffractive scattering at given $b$.

Coefficient K is gotten to take into account  that probability of producing signal object with 'humps' at resolved $\eta$ range is not equal to the one without 'humps'. $K$ usually can be calculated because of the hardness of the signal object, for the higgs at LHC case, $K$ is about \cite{SGP1} 10.

Strictly speaking, we can calculate $P^{SO}_{white}$ from any of the equations (\ref{eqB}),(\ref{eqC}), if we know all other quantities, but at reality we don`t know $P^{SO}_{color}$. So, we must exclude $P^{SO}_{color}$ from equations (\ref{eqB}),(\ref{eqC}) to calculate $P^{SO}_{white}$:
\begin{equation}
P^{SO}_{white}=\frac{P_{C}-P_{DD}P_{B}}{P_{inelastic}\left(K-P_{DD}\right)}
\label{eqSOw}
\end{equation}
In general case, equations  (\ref{eqB}),(\ref{eqC}) is non-linear and more complex, but they can be solved to get $P^{SO}_{white}$ without knowing $P^{SO}_{color}$.

We have to mention, that equation (\ref{eqSOw}) (or its generalization in real case) gives us possibility to determine $P^{SO}_{white}$ even if survival gap probability is zero,  in the absence  of the straightforward process, shown as  A on Fig.\ref{fig:figABC}.

It was shown \cite{SGP1}, that survival  probability for central rapidity gap in the higgs production is low, about $0.01 \div 0.15$, and, so, our method can be applied.

Let`s make brute estimation of applicability of our method to higgs production. The most natural way to detect higgs and to determine higgs mass is to observe differential cross-section $\frac{d\sigma}{dM}$, where $M$ is the mass of the system high-energetic products of higgs decaying, such as leptons for leptonic decaying modes or b-jets for $H \rightarrow b\overline{b}$ decaying mode. To estimate these cross-sections we can assume, that form of profiles of all probabilities at (\ref{eqB}),(\ref{eqC}) is the same, and we can integrate that equations in $b$. 

\begin{equation}
\frac{d\sigma_A}{dM}=\frac{d\sigma^{H}_{WW}}{dM}S^2
\label{eqA1}
\end{equation}
\begin{equation}
\frac{d\sigma_B}{dM}=\frac{d\sigma^{SO}_{color}}{dM}+\frac{d\sigma^{H}_{WW}}{dM}(1-S^2)
\label{eqB1}
\end{equation}
\begin{equation}
\frac{d\sigma_C}{dM}=\frac{d\sigma^{SO}_{color}}{dM}\frac{\sigma_{DD}}{\sigma_{tot}}+K\frac{d\sigma^{H}_{WW}}{dM}(1-S^2)
\label{eqC1}
\end{equation}

Value of $\frac{d\sigma^{SO}_{color}}{dM}$ is process-specific, it is defined  by higgs decaying channel and by final-state selection procedure. This background cross-section can be estimated as the sum of the background and signal cross sections for the $gg \rightarrow H$ channel. First one is much larger than second one, so, we can assume that background  $\frac{d\sigma^{SO}_{color}}{dM}$ have the same value as the background for $gg \rightarrow H$ channel.

First addendum in (\ref{eqB1}) is much larger, than one in (\ref{eqC1}), but second addendum in (\ref{eqB1}) is much smaller, than one in (\ref{eqC1}).

\begin{figure}[th]
\centerline{\psfig{file=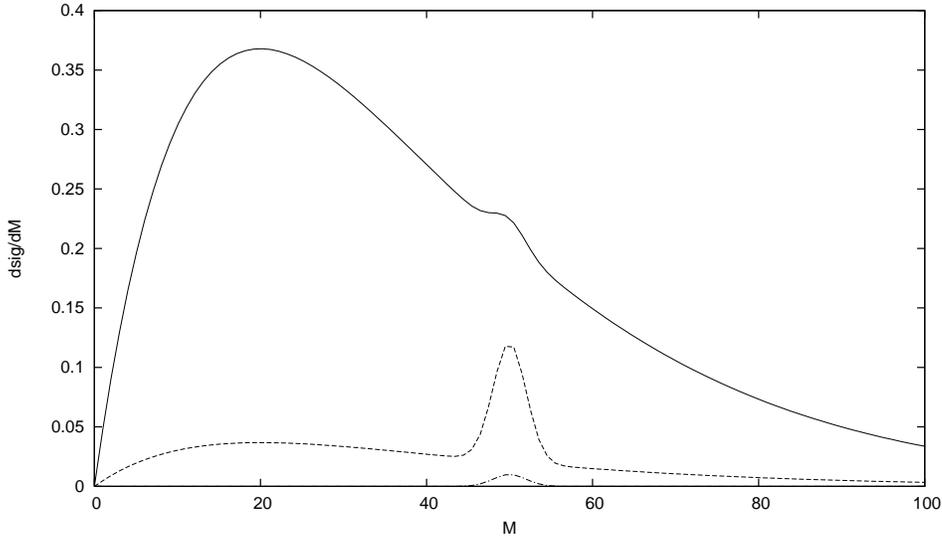,width=5in}}
\vspace*{0pt}
\caption{Estimated cross-sections $\frac{d\sigma}{dM}$ for the signal events.  Upper solid  curve is for events, then signal object with mass $M$ detected and no 'humps' in resolved $\eta$ range is presented. Middle dashed curve is for events, then signal object with mass $M$ detected and two 'humps' on the both sides of the signal object is presented. In both cases, there are the plateau of soft particles  on the whole $\eta$ range. Lower dot-dashed curve is for events with signal object with rapidity gaps on both sides.}
\label{fig:figdsigdm}
\end{figure}

Expected behavior for $\frac{d\sigma}{dM}$ for examined types of events is schematically drawn on Fig.\ref{fig:figdsigdm}.

Direct way to detect higgs from this cross-sections is to multiply $\frac{d\sigma_B}{dM}$  (upper curve on Fig.\ref{fig:figdsigdm}) by the factor $\frac{\sigma_{DD}}{\sigma_{tot}}$ and to substitute it from $\frac{d\sigma_C}{dM}$  (middle curve on Fig.\ref{fig:figdsigdm}). If there is no weak boson fusion mechanism of higgs production, result will be zero. In other case, we will get lower curve on Fig.\ref{fig:figdsigdm}, multiplied by the factor $\frac{1}{S^2} \sim 10$.
 
 Practically, this method can be applied as a part of higgs trigger. We can extract from the experiment avaerage number of particles inside of rapidity region of the two tag jets and outside of them:
 \begin{equation}
 \begin{array}{l}
 n_{inside}=\frac{N_{inside}}{\Delta\eta} \\
 n_{outside}=\frac{N_{outside}}{\eta_{max}-\eta_{min}-\Delta\eta} \\
 \end{array}
 \end{equation}
 there $\Delta\eta$ is the rapidity distance between tag jets.
 
As we`ve shown above, this quantities must be approximatly equal in the case of the trivial colour channel and must be different in the case of the signal colour-less channel. So, the best quantity to observe is
\begin{equation}
\Delta n=n_{outside}-n_{inside}
\end{equation}
This quantity can be used in the modern-state cut-to-cut anlisys by choosing some critical value $\Delta n_{cut}$. If $\Delta n$ is greater than  $\Delta n_{cut}$, event is accepted to be signal and rejected otherwise. In the liklyhood analysys this value can be used too (and it is more preferable).

Let`s discuss advantages and lacks of this advanced gap method.
 
Proposed type of events is a half-way between $gg \rightarrow H$ channel and $WW\rightarrow H$ with rapidity gap channel. As compared with gluon fusion channel, we have suppressed by the factor $\frac{\sigma_{DD}}{\sigma_{tot}}$ background and suppressed by the factor $\frac{\sigma(WW\rightarrow H)}{\sigma(gg\rightarrow H)}$ signal. As compared with weak boson fusion with rapidity gap method, we have increased the signal by the factor $\frac{1}{S^2}$ and have add some substantial background.
 
Another advantage of our method is possibility of cross-checking, because we investigate all three type of events with only two unknown cross-sections, signal  $\frac{d\sigma^{H}_{WW}}{dM}$ and background $\frac{d\sigma^{SO}_{color}}{dM}$.
 
 Lacks of our method can be divided to two classes.
 
At first, we add statistical uncertainty, because of 'humps' on plateau can be generated by statistical fluctuations of $\frac{dn}{d\eta}$. This factor can be easily calculated, but we can not remove this uncertainty. 
 
At second, we have theoretical uncertainty in the soft interactions. We don`t know any reliable way to calculate  $P_{inelastic}$ and $P_{DD}$ in equations (\ref{eqB}),(\ref{eqC}) and we don`t know, is the probabilities in these equations factorizable or not. This uncertainty can be removed, if we will construct reliable theory of the soft (Pomeron) interactions. We can generalize this problem as the problem of constraction of the 'soft' generator.
 
\section*{Acknowledgments}
We thank N.Prikhod`ko for useful discussions.
This work was supported by RFBR Grant RFBR-03-02-16157a, grant of Ministry for Education E02-3.1-282 and St.Petersburg grant í04-2.4ë-364.

\end{document}